\pdfoutput=1
\documentclass[showpacs,amsmath,amssymb,aps,pre,twocolumn]{revtex4-1}
\usepackage{graphicx}
\usepackage{dcolumn}
\usepackage{bm}
\usepackage[breaklinks,colorlinks = true,linkcolor = red,urlcolor=blue,citecolor=red]{hyperref}
\usepackage{multirow}
\usepackage{array}
\usepackage{booktabs}
\usepackage{ctable}
\usepackage{ulem}
\usepackage{upgreek}
\usepackage{epsfig}
\usepackage{mathrsfs}
\usepackage{amssymb}
\usepackage{amsbsy}
\usepackage{color}
\usepackage{cancel}
\usepackage{pifont}
\usepackage{marginnote}
\usepackage{float}
\usepackage{verbatim}
\usepackage{subfigure}
\usepackage{bbm, dsfont}
\usepackage{lineno}
\usepackage{amsmath}
\definecolor{dgreen}{rgb}{0,0.7,0}

\def\bea{\begin{eqnarray}}
\def\eea{\end{eqnarray}}

\def\nn{\nonumber}

\makeatletter

\newcommand{\Rmnum}[1]{\expandafter\@slowromancap\romannumeral #1@}
\makeatother

\newcommand{\eref}[1]{Eq.~(\ref{#1})}%
\newcommand{\fref}[1]{Fig.~\ref{#1}} %
\newcommand{\aref}[1]{Appendix~\ref{#1}}%
\usepackage[utf8]{inputenc}

\begin{abstract}
{What determines the average length of a queue which stretches in front of a service station? The answer to this question clearly depends on the average rate at which jobs arrive to the queue and on the average rate of service. Somewhat less obvious is the fact that stochastic fluctuations in service and arrival times are also important, and that these are a major source of backlogs and delays. Strategies that could mitigate fluctuations induced delays are in high demand as queue structures appear in various natural and man-made systems. Here we demonstrate that a simple service resetting mechanism can reverse the deleterious effects of large fluctuations in service times, thus turning a marked drawback into a favourable advantage. This happens when stochastic fluctuations are intrinsic to the server, and we show that the added feature of service resetting can then dramatically cut down average queue lengths and waiting times. While the analysis presented herein is based on the M/G/1 queueing model where service is general but arrivals are assumed to be Markovian, Kingman's formula asserts that the benefits of service resetting will carry over to queues with general arrivals. We thus expect results coming from this work to find widespread application to queueing systems ranging from telecommunications, via computing, and all the way to molecular queues that emerge in enzymatic and metabolic cycles of living organisms.}
\end{abstract}

\begin{document}


\title{Mitigating long queues and waiting times with service resetting}

\author{Ofek Lauber Bonomo$^{1}$, Arnab Pal$^{2}$  and Shlomi Reuveni$^{1}$}

\affiliation{$^{1}$School of Chemistry, Raymond and Beverly Sackler Faculty of Exact Sciences \& The Center for Physics and Chemistry of Living Systems \&   The Ratner Center for Single Molecule Science, Tel Aviv University, Tel Aviv 6997801, Israel} 

\affiliation{$^{2}$Department of Physics, Indian Institute of Technology, Kanpur, Kanpur 208016, India} 

\date{\today}

\maketitle

\section{Introduction}
\label{Introduction}
\begin{figure}[t!]
\includegraphics[scale=0.27]{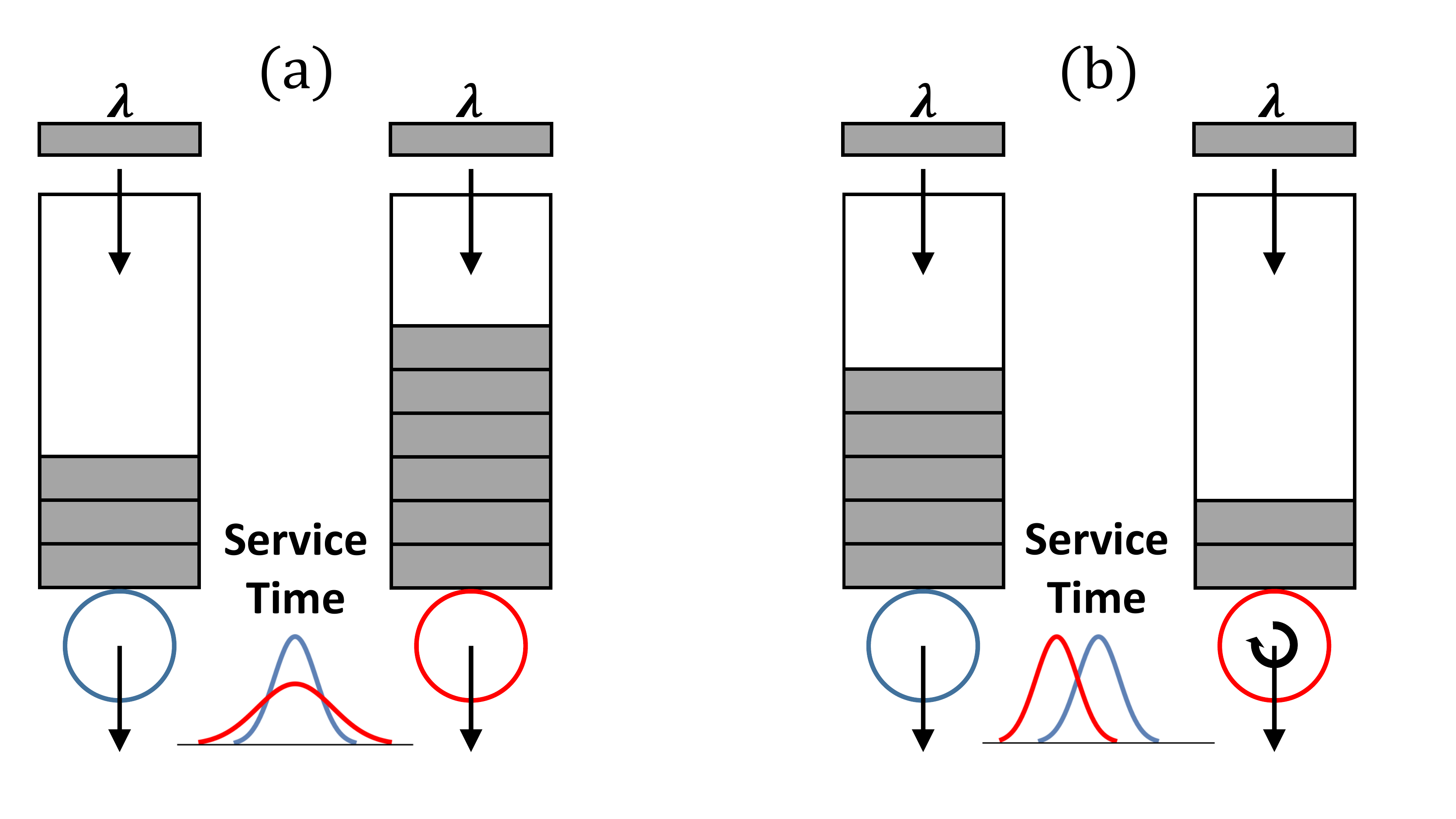}
\caption{Backlogs and delays caused by service time fluctuations can be mitigated with service resetting. Panel (a): The mean number of agents---jobs, customers, molecules, etc.---in a queue is sensitive to stochastic fluctuations in the service time provided by the server (computer, cashier, enzyme, etc.). For a given arrival rate and mean service time, the queue stretching in front of the server whose service time distribution is more variable will be longer on average, as illustrated by the blue and red servers in the figure. Panel (b): We show that when large service time fluctuations are intrinsic to the server, resetting the service process can lower both the mean and variance of the service time and drastically improve queue performance.}
\label{fig1}
\end{figure}

Queueing theory is the mathematical study of waiting lines \cite{Cohen-book,Newell-book,Adan-book,Haviv-book}. Ranging from the all familiar supermarket and bank, to call centers \cite{Call1, Call2}, airplane boarding \cite{Plane1, Plane2, Plane3},  telecommunication and computer systems \cite{CSqueue1,Tele1,CSqueue2,Mor-book,Nelson-book,Tele2}, production lines and manufacturing \cite{PL1,PL2,Askin-Book,Curry-Book}, enzymatic and metabolic pathways \cite{enzymatic1,enzymatic2,enzymatic3,enzymatic4,enzymatic5,enzymatic6,enzymatic7,enzymatic8}, gene expression \cite{Gene1,Gene2,Gene3,Gene4,Gene5,Gene6,Gene7}, and in transport phenomena \cite{Transport1,Transport2,Transport3,Transport4,Transport5,Transport6,Transport7,Transport8,Transport9,Transport10,Transport11,Transport12}, waiting lines and queues appear ubiquitously and play a central role in our lives. While the teller at the bank works at a (roughly) constant rate, other servers, e.g., computer systems \cite{Mor-book}, and molecular machines like enzymes \cite{Large_fluc1,Large_fluc2,Large_fluc3,Large_fluc4,Large_fluc5,Large_fluc6}, often display more pronounced fluctuations in service times. These fluctuations have a significant effect on queue performance \cite{Cohen-book,Newell-book,Adan-book,Haviv-book}:  
higher fluctuations in service times will result in longer queues as illustrated in Fig. \ref{fig1}a. Service time variability is thus a major source of backlogs and delays in queues \cite{Mor-book}, and this problem is particularly acute when encountering heavy tailed workloads \cite{Whitt-2000}. 

Different strategies have been developed to mitigate the detrimental effect caused by stochastic service time fluctuations. In particular, when considering single server queues, various scheduling policies can be applied to reduce waiting times. For example, computer servers can be designed to serve smaller jobs first, rather than by order of arrival. While this policy can be criticized from the standpoint of fairness, it does reduce average waiting times by preventing situations where typically sized jobs (which are common) get stuck behind a single job that is very large \cite{Mor-book}. In situations where service can be stopped and continued from the same point at a later time, one can further improve performance by implementing a policy in which only the job with the shortest remaining service time is served \cite{SRPT}. This policy can be proven optimal under certain conditions \cite{SRPT-OPT}.

The problem with the above-mentioned scheduling policies is that they are ill-equipped to deal with situations where fluctuations in service times are intrinsic to the serves itself. These are in fact prevalent. For example, stochastic optimization algorithms can take wildly different times to solve two instances of the \textit{exact same} problem \cite{SO}. Similarly, the time it takes an enzyme to catalyze a given chemical reaction varies considerably between turnover cycles --- despite the fact that the incoming (substrate) and outgoing (product) molecules are chemically \textit{identical} \cite{Large_fluc1,Large_fluc2,Large_fluc3,Large_fluc4,Large_fluc5,Large_fluc6}. Crucially, in these and similar cases the total service time of an incoming job is unknown a priori and the time remaining for a job in service cannot be determined. Sized-based scheduling policies are thus impossible to implement, which calls for the development of new and novel approaches to the problem.

In this paper we propose a complementary approach to solve the problems caused by service time fluctuations in queueing systems. Namely, we show that implementing a simple service resetting policy can reverse the deleterious effect of stochastic fluctuations when the latter are large and particularly harmful. The policy, which consists of random or deterministic resetting of the service process, may seem counterintuitive at first. After all, no benefit can possibly come from doing the exact same thing all over again. However, here we show that when service time fluctuations are intrinsic to the server itself, in the sense that jobs whose service has been reset are assigned fresh service times, service resetting can be used to drastically improve queue performance as illustrated in Fig. \ref{fig1}b.  
Indeed, it has recently been shown that resetting has the ability to expedite the completion of random processes: from stochastic optimization \cite{CS1,CS2,CS3,CS4}, via first-passage and search \cite{Restart1,Restart2,branching,Review,ReuveniPRL16,PalReuveniPRL17, Restart-Search1,Restart-Search2, Restart-Search3, interval,interval-v,Peclet,Das1,Das2,expt,expt2,expt3}, and onto chemical reactions \cite{ReuveniEnzyme1,ReuveniEnzyme2,ReuveniEnzyme3}, and this general principle is hereby exploited in the context of queueing. 

A different way in which resetting can be used to affect queue performance is by resetting of the jobs arrival process, which is interesting to consider e.g., in the context of intracellular transport \cite{queue-input}. However, resetting the arrival process has no affect on service time fluctuations, which is further compounded by the fact that in most conventional queue structures the arrival process is extrinsic to the system and is thus not subject to control or optimization. We thus focus on queues with service resetting to which we devote this paper. Before moving forward to develop the theory and discuss examples, we mention in passing that the concept of resetting is also relevant in the context of queues with catastrophic events  \cite{Q-w-C-1,Q-w-C-2,Q-w-C-3,Q-w-C-4,Q-w-C-5,Q-w-C-6,Q-w-C-7,Q-w-C-8,Q-w-C-9,Q-w-C-10,Q-w-C-11,Q-w-C-12,Q-w-C-13,Q-w-C-14,Q-w-C-15}. 
Note, however, that such catastrophic events  lead to mass annihilation of jobs (agents) from the queue. Contrary, service resetting conserves the number of jobs in the queue, as jobs whose service was reset would still require full service before they can leave the system. Thus, `resetting by catastrophe' should not to be confused with the service resetting policy that is introduced and  rigorously analyzed below.

The remainder of this paper is structured as follows. In Sec. \ref{sec 2}, we provide a brief review of the M/G/1 queuing model for which queue arrivals are Markovian and service is general. This queue will serve as the main modeling platform of this paper. We also review the Pollaczek-Khinchin formula which provides the mean number of jobs in the M/G/1 queue at the steady-state, highlighting the sensitivity of the latter to service time fluctuations. Finally we provide a short overview of existing scheduling strategies aimed to mitigate service time fluctuations. In Sec. \ref{Sec 3}, we discuss the service resetting policy and formulate the M/G/1 model with service resetting. We show the latter can be mapped onto the standard M/G/1 queue, but with a modified service time distribution that depends on the underlying distributions of the service and resetting times. This fact allows us to re-apply the Pollaczek-Khinchin formula to study how the mean number of jobs in the queue depends on the resetting protocol. We focus on two prominent resetting protocols for which we provide detailed analysis: In Sec. \ref{poisson resetting sec} we consider Poissonian resetting. i.e., resetting at a constant rate. We show that when the variability in the underlying service time distribution is high, the  mean number of jobs in the system obtains a minimum as a function of the resetting rate. This means that resetting can drastically improve queue performance. Similarly, in Sec. \ref{sharp resetting sec} we study resetting at fixed time intervals (a.k.a sharp resetting), and show that while this resetting protocol leads to similar qualitative results, the fixed inter-resetting time duration can be tuned to perform better than any other stochastic resetting protocol (Poissonian resetting included). In Sec. \ref{examples} we illustrate the general results obtained for two well-established service time distributions, namely the log-normal and Pareto distribution, which are considered respectively in subsections \ref{Log-normal service} and \ref{Pareto service}. We end in Sec. \ref{Conc} with conclusions and outlook. 

In what follows we use $f_Z(t)$, $q_Z(t)=1-\int_0^t~d\tau~f_Z(\tau)$,  $\langle Z \rangle$, Var$(Z)$, and $\tilde{Z}(s)~\equiv~\langle e^{-sZ}\rangle$ to denote, respectively, the probability density function, the survival function, expectation, variance, and Laplace transform of a non-negative random variable $Z$. 

\section{Queues: models and preliminaries} \label{sec 2}
Consider a queuing system that is composed of a queue in which jobs await to be served, and a server which serves one job at a time according to a First-Come, First-Served policy (FCFS).
A queue can be identified as a stochastic process whose state space is denoted by the set $N=\{ 0,1,2,3,... \}$,
where the value corresponds to the number of jobs in the queue, including the one 
being served. In particular,
an M/G/1 queue (written in Kendall's notation \cite{Kendall's-notation}) consists of a single server and queue to which jobs arrive according to a Poisson process with rate $\lambda$ \cite{Nelson-book,Mor-book}. Thus, M stands for Markovian or memory-less arrival process. No assumptions are made on the service time of jobs, which comes from a general distribution (hence the G in Kendall's notation). Denoting the service time by the random variable $S$ (for service), whose density we denote by $f_S(t)$, we define the service rate $\mu \equiv 1/\langle S \rangle$. It will prove convenient to introduce the utilization parameter $\rho=\lambda/\mu$, which gives the fraction of time the server is working (not idle) in the steady state \cite{Mor-book}. The model attains a steady state as long as the arrival rate is smaller than the service rate, i.e, $\rho<1$. For $\rho>1$ the queue grows indefinitely long with time and the system does not attain a steady state.  Thus, in what follows we will assume $\rho<1$. 


The number of jobs in an M/G/1 queue fluctuates with time. Thus, it is only natural to start the discussion with the average of this observable. The mean number of jobs $\langle N \rangle$ 
in the system (queue+server) is given by the famous Pollaczek-Khinchin formula \cite{Mor-book}
\bea
\langle N \rangle
=\frac{\rho}{1-\rho}+\frac{\rho^2}{2(1-\rho)}\left( CV^2-1  \right)~,
\label{PK-1}
\eea
where $\rho$ is the utilization, and $CV^2=\text{Var}(S)/\langle S \rangle^2$ is the squared coefficient of variation or the variability in service time.
As expected, the mean number of jobs is a monotonically increasing function of the utilization, with $\langle N \rangle$ tending to infinity as $\rho\rightarrow 1^{-}$. Thus, as expected, low service rates lead to long queues. Note, however, the appearance of $CV^2$ in the second term of \eref{PK-1}. This indicates that the mean length of the queue is highly sensitive to stochastic fluctuations in the service time, which is a non-trivial effect that can be explained by the inspection paradox \cite{Mor-book}.

From the second term in \eref{PK-1}, we see that the mean number of jobs is a monotonically increasing function of $CV^2$. Note, that when $CV<1$, i.e., when service time fluctuations are relatively small, the contribution of the second term is negative, leading to shorter queues. On the other hand, when $CV>1$, i.e., when service time fluctuations are relatively large, the contribution of the second term is positive, leading to longer queues. Importantly, there can be a ``piling up'' of jobs due to high service time variability. This can happen, for example, if short service times are occasionally followed by extremely-long service times. In such situations, $CV^2$ is large, resulting in long queues, even in low utilization. It is thus apparent that fluctuations in the service time play a central role in the behavior of M/G/1 queues.

Finally, we recall that the mean waiting time $\langle T \rangle$ of a job in the queue, i.e., the time elapsed from arrival to the end of service, follows from \eref{PK-1} by using Little's law \cite{Mor-book} which states $\langle T \rangle=\lambda^{-1}~\langle N \rangle$. Thus, we have
\bea
\langle T \rangle=
\frac{1/\mu}{1-\rho}+\frac{\rho/\mu}{2(1-\rho)}\left( CV^2-1  \right)~,
\label{PK-2}
\eea
and note that similar to the mean number of jobs, the mean waiting time crucially depends on fluctuations in service time.

To deal with the detrimental effect caused by service time fluctuations several different strategies have been developed. When considering single server queues, the main tools at our disposal are scheduling policies (a.k.a service policies) \cite{SchedPol}. A scheduling policy is the rule according to which jobs are served in a queue. For example, servers can serve jobs according to the order of their arrival as e.g. happens in the supermarket. This simple First-Come, First-Served policy (FCFS) seems fair when working with people as customers. However, when considering queues of inanimate objects, fairness is not necessarily the most important  requirement and other service policies can also be considered.

The FCFS policy is an example of a \textit{non-sized-based policy} --- the server serves jobs based on their order of arrivals rather than their size. However, in situations where jobs sizes are known, one can utilize this knowledge to devise size-based policies such as Shortest-Job-First (SJF). According to this policy, once free, the server chooses to work on the job with the smallest size and hence also the shortest service time. In terms of the average waiting time of a job in the queue, the SJF policy performs better than the FCFS policy \cite{Mor-book}. This can be understood intuitively. In the FCFS policy, all jobs that arrive after a very large job will suffer from a long waiting time. This will not happen in the SJF policy, where extremely large jobs which are rare will not be selected for service before jobs of typical size. The latter are much more common and since they are served first they spend less time in the queue, resulting in smaller waiting times on average compared to \eref{PK-2}. 

In the SJF policy, the server chooses the smallest job waiting in the queue each time service is complete. This approach can be further developed allowing the server to inspect jobs' sizes (including the one in service) continuously in time---constantly searching for the job with the \textit{shortest remaining service time}. In such a queue, an arriving job can preempt the job in service if its own service time is lower. The service of the preempted job is resumed later, starting from the point where it was stopped. The above policy is called Shortest-Remaining-Processing-Time (SRPT) \cite{SRPT}, and it was used e.g in web servers to reduce the waiting time of static HTTP requests whose sizes have been shown to follow a heavy-tailed distribution \cite{q-control-3,q-control-4}. It was proved in \cite{SRPT-OPT}, that in single-server queues where the service time of jobs is known and preemption is possible, the SRPT policy is the optimal policy with respect to minimizing the mean waiting time.

Having readily available knowledge of job sizes and their remaining service times, as well as the ability to preempt jobs, is a luxury not shared by all queueing systems. For example, in non-deterministic algorithms, e.g., stochastic optimization methods, inputs of similar size can render significantly different run times \cite{SO}, and the same can happen for consecutive runs of the algorithm using the same input. Sized-based scheduling policies like the SJF and SRPT are then inapplicable, but we will hereby show that the intrinsic randomness of the service process can be exploited via service resetting to achieve similar performance goals.

\section{M/G/1 Queues with service resetting}
\label{Sec 3}
In Sec. \ref{sec 2} we reviewed the non-trivial dependence, of the mean number of jobs in an M/G/1 queue, on service time fluctuations. Namely, large relative fluctuations lead to long queues. In this section, we analyze the effect of service resetting on the mean and variance of the service time.  

To understand how restarting service affects the overall service time of a job, we consider two extreme scenarios. First, consider a situation where fluctuations in service times are \textit{extrinsic} to the server. Thus, imagine a server that serves jobs of variable size at a constant rate. In this case, the service time is determined exclusively by the job size, i.e., the bigger the job the longer the service time, and vice versa. An example of such would be a supermarket cashier counter. The teller serves the customers at a (roughly) constant rate, and the service time is determined by the number of items each customer has. Thus, the fluctuations in service time originate solely from the variability in the number of items each customer brought. Imagine now that the teller decides to restart service from time to time. Since restart does not affect the number of items one has, this strategy is clearly detrimental. The time already spent in service is lost, while the customer's required service time remains unaltered. Thus, restart results in wasted time and delays. 
  
Now, consider a queue in which fluctuations in service times are \textit{intrinsic} to the server, i.e., a queue in which all jobs are (roughly) identical but the time it takes to serve a job is nevertheless random. 
Example of such would be a computer server which runs a stochastic algorithm. Such algorithm employs probabilistic approaches to the solution of mathematical problems, e.g., optimization. The run time of such an algorithm can vary considerably between runs. Importantly, stochastic fluctuations in run times come from the probabilistic solution method, and would hence differ even between identical instances of the same problem. Thus, resetting such an algorithm in its course of action would result in a new and random run time, contrary to the case of a teller in a supermarket. 

Another example of a queue in which fluctuations in service time are intrinsic to the server can be found in enzymatic reactions. In the context of enzymes, substrate molecules can be viewed as costumers, forming a `waiting line' to the enzyme which acts as a server. Substrate molecules are identical, and require the same type of service: a catalytic process which converts a substrate molecule to a product molecule. Yet, at the single molecule level, chemistry is stochastic as thermal fluctuations render the service (catalysis) time random. It is often the case that a substrate molecule unbinds the enzyme without completing service \cite{ReuveniEnzyme1}. In this case, service is reset, and a new and random service time is drawn upon rebinding.

Motivated by the examples above, we will now consider queues in which service time fluctuations are intrinsic to the server. As explained above, in these queues resetting results in a newly drawn service time that is added to the time already spent in service. In what follows, we will quantify the effect of resetting on the mean and variance of the total service time. We will then go on to show that in certain situations resetting significantly expedites service, thus improving overall performance by shortening queues. This effect will be discussed in the next section.

Let us consider an M/G/1 queue with service resetting. Assume that both the service and restart times are two independent and generally distributed random variables. The total service time under restart, $S_R$, is then described by the following renewal equation 
\begin{equation}
\begin{array}{l}
S_{R}=\left\{ \begin{array}{lll}
S &  & \text{if ~~}S<R\text{ ,}\\
 & \text{ \ \ }\\
R+S_R'&  & \text{if~~ }R\leq S\text{ ,}
\end{array}\right.\text{ }\end{array}
\label{renewal-1-main}
\end{equation}
where $R$ is the random resetting time drawn from a distribution with density $f_R(t)$, and $S_{R}'$ is an independent and identically distributed copy of $S_{R}$. To understand this equation, observe that when service occurs before restart, $S_R = S$. However, if service is restarted at a time $R \leq S$, then a new service time is drawn, and service starts over. In this case, $S_R = R + S_R'$.

Service can be seen as a first passage process which ends when a job is served. A comprehensive framework for first passage under restart was developed in \cite{ReuveniPRL16,PalReuveniPRL17}. In the following, we show how the results obtained there can be used to gain insight on the performance of an M/G/1 queue under restart. Starting from \eref{renewal-1-main}, one can obtain the probability density of $S_R$ in Laplace space (\aref{SR-PDF-LT}). Here, we will only be interested in the first two moments, which are given by (\aref{SR-PDF-LT})
\begin{align}
\langle S_R  \rangle &= \frac{\langle \text{min}(S,R) \rangle}{\text{Pr}(S<R)} ,
\label{first-mom} \\
\langle S_R^2  \rangle &= \frac{\langle \text{min}(S,R)^2 \rangle}{\text{Pr}(S<R)}+
\frac{2 \text{Pr}(R \leq S) \langle   R_{\text{min}}\rangle \langle \text{min}(S,R) \rangle}{\text{Pr}(S<R)^2},
\label{sec-mom}
\end{align}
where $\text{min}(S,R)$ is the minimum between $S$ and $R$, $\text{Pr}(S<R)$ is the probability of service being completed prior to restart, and $R_{\text{min}}=\{R|R=\text{min}(R,S)\}$
standing for the random restart time given that restart occurred prior to service. Finally, recall that the variance in the service time is given by $\text{Var}(S_R)=\langle S_R^2 \rangle -\langle S_R \rangle ^2$ which will be useful in the next section. Equations (\ref{first-mom}) \& (\ref{sec-mom}) assert that the mean and variance of the service time under restart can be evaluated directly from the distribution of the resetting time and the distribution of the service time without resetting. Importantly, in cases where this cannot be done analytically, numerical methods can be used to obtain $\langle S_R  \rangle$ and $\langle S_R^2  \rangle$ and as long as the distributions of $R$ and $S$ are known or can be sampled from.

\section{M/G/1 Queues with Poissonian service resetting} \label{poisson resetting sec}
Poissonian resetting, i.e., resetting with a constant rate, has been extensively investigated  \cite{Restart1,Restart2,Restart-Search1,Restart-Search2,branching,ReuveniPRL16,PalReuveniPRL17, interval,interval-v,Peclet,Das1,Das2,expt,expt2,expt3, Restart-Search3,Pal-potential,GBM,EVS-1,EVS-2} (see \cite{Review} for extensive review and \cite{IP} for discussion on the connection with the inspection paradox). As the name suggests, here resetting follows a Poisson process and the number of resetting events in a given time interval comes from the Poisson distribution. In this section we quantify the effect of Poissonian resetting on the mean and variance of the total service time and the mean length of an M/G/1 queue.

Consider an M/G/1 queue with a server that is restarted at a rate $r$. In other words, we take the restart time $R$ to be an exponential random variable with mean $1/r$, and restart is thus a Poisson process with rate $r$. The mean and second moment of the service time can then be derived using Eqs. (\ref{first-mom}) and (\ref{sec-mom}), giving  (\aref{A1})
\bea
\langle S_r \rangle &=& \frac{1-\tilde{S}(r)}{r \tilde{S}(r)}, \label{mean under restart} \\
\langle S_r^2 \rangle &=& \frac{2\left( r \frac{d \tilde{S}(r)}{dr}-\tilde{S}(r) +1 \right)}{r^2 \tilde{S}(r)^2},
\label{secmom under restart}
\eea
where $\tilde{S}(r)=\int_0^\infty~dt~ e^{-rt}~f_S(t)$ is the Laplace transform of the service time, evaluated at the restart rate $r$.

The utilization of this queue is then given by $\rho_r=\lambda\langle S_r \rangle$, and the squared coefficient of variation of the service time is  $CV_r^2=\text{Var}(S_r)/\langle S_r \rangle^2$. We can now write the mean queue length under restart by replacing $\rho$ by $\rho_r$, and $CV^2$ by $CV_r^2$, in \eref{PK-1}. This yields  
\bea
\langle N_r \rangle=\frac{\rho_r}{1-\rho_r}+\frac{\rho_r^2}{2(1-\rho_r)}\left( CV_r^2-1  \right)~.
\label{PK-1-r}
\eea
Similarly, the mean waiting time $\langle T_r \rangle$ in the system can be derived from Little's law \cite{Nelson-book,Mor-book}, yielding an analogous result to \eref{PK-2}.

To better understand the effect of resetting on the mean queue length, consider the introduction of an infinitesimal resetting rate $\delta r$. Utilizing \eref{mean under restart}, we then find 
\bea 
\langle S_{\delta r} \rangle \simeq \langle S \rangle-\delta r \frac{\langle S \rangle^2}{2} \left[ CV^2-1  \right]+\mathcal{O}(\delta r^2). \label{mean service time expansion}
\eea 
As expected, the first term on the right hand side of \eref{mean service time expansion} is the mean of the \textit{original} service time, i.e., without resetting.
The second term gives the first order correction, and note that its sign is governed by $CV^2$ of the \textit{original} service time. Specifically, for $CV^2>1$, we have $\langle S_{\delta r} \rangle < \langle S \rangle $ and vice versa. In other words, the introduction of Poissonian resetting to an M/G/1 queue will reduce the mean service time whenever the coefficient of variation of the original service time is greater than unity.

\begin{figure}[t]
\includegraphics[scale=0.9]{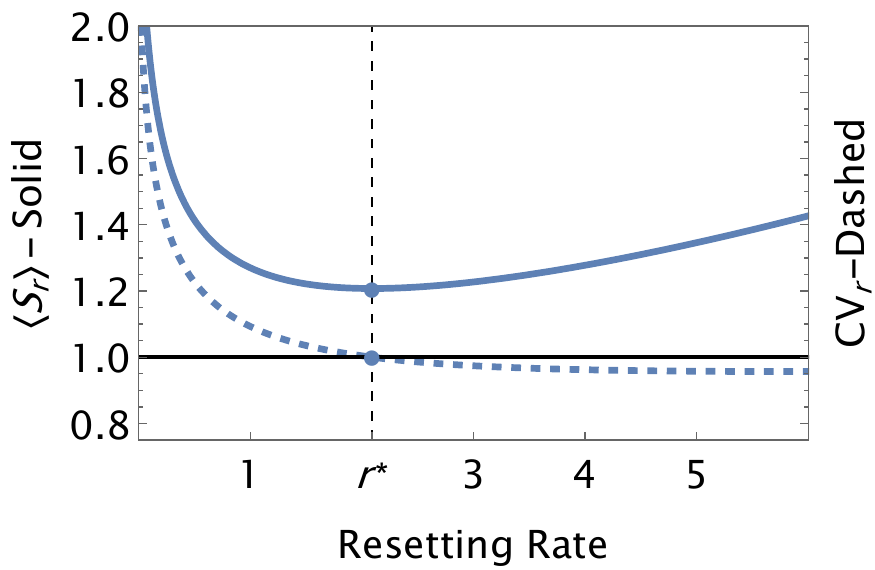}
\caption{The mean (solid line) and $CV$ (dashed line) of the service time with Poissonian resetting as a function of the resetting rate. Plots were made, using Eqs. (\ref{mean under restart}) and (\ref{secmom under restart}), for an underlying service time taken from the inverse Gaussian distribution whose density is given by $f_S(t) = \sqrt{\gamma / 2\pi t^3} e^{-\gamma (t-\mu)^2 / 2\mu ^2 t},~t>0$. Here, $\mu = 2.5$ and $\gamma = 0.5$. Observe that the mean service time with resetting, $\langle S_r \rangle$, is minimized at an optimal resetting rate $r^* \simeq 2.092$ which was found using \eref{r* equation}. At this optimal resetting rate we have  $CV_{r^*}=1$.}
\label{fig2}
\end{figure}

In Fig. \ref{fig2}, we consider a representative example for the effect of resetting on an M/G/1 queue with large fluctuations in service time. Starting from a service time distribution with $CV>1$, we introduce resetting and plot the mean service time $\langle S_r \rangle$ vs. the resetting rate. As predicted by \eref{mean service time expansion}, we see that $\langle S_r \rangle$ initially decreases, obtaining a minimum at an optimal resetting rate $r^*$. In addition, note that at this optimal resetting rate we have  $CV_{r^*}=1$. Remarkably, this observation is not specific to the service time distribution considered in Fig. \ref{fig2}, but is rather a universal property. Namely, it can be shown that \cite{PalReuveniPRL17}

\bea 
CV_{r^*}=1, \label{Universal relation CV}
\eea 
for any resetting rate,  $0<r^*<\infty$, at which
\bea
\frac{d \langle S_r\rangle}{dr}\bigg|_{r^*}=0. \label{Exp resetting opt}
\eea
Substituting Eq. (\ref{mean under restart}) into Eq. (\ref{Exp resetting opt}) yields 
\bea
\tilde{S}^2(r^*)-\tilde{S}(r^*)-r^*\tilde{S}'(r^*)=0, \label{r* equation}
\eea
which can be solved to obtain $r^*$. Combining \eref{Universal relation CV} with the Pollaczek-Khinchin formula given in \eref{PK-1-r}, we have
\bea
\langle N_{r^*} \rangle=\frac{\rho_{r^*}}{1-\rho_{r^*}}.
\label{PK-1-r*}
\eea
Note that the mean number of jobs at optimality is equivalent to the mean number of jobs in an M/M/1 queue \cite{Mor-book}, where job service times are exponentially distributed with mean $\langle S_{r^*} \rangle$.    

Comparing Eqs. (\ref{PK-1}) and (\ref{PK-1-r*}), we observe that $\langle N \rangle > \langle N_{r^*} \rangle$. One can easily derive this inequality by substituting $CV>1$ into \eref{PK-1}. This yields 
\bea 
\langle N \rangle &=& \frac{\rho}{1-\rho}+\frac{\rho^2}{2(1-\rho)}~(CV^2-1) > \frac{\rho}{1-\rho} \nn \\
&>& \frac{\rho_{r^*}}{1-\rho_{r^*}} = \langle N_{r^*} \rangle, \label{Nr* optimality}
\eea 
where in the last inequality we used the monotonicity of the function $\frac{\rho}{1-\rho}$ on the interval $[0,1)$, and the optimality of $r^*$, which implies $\langle S_{r^*}\rangle< \langle S\rangle$ resulting in $\rho_{r^*}< \rho$.

We see that whenever $CV>1$ for the original service time, the mean number of jobs in the queue can be reduced by resetting service at an optimal rate. When doing so, one also sets the coefficient of variation of the optimally restarted service time to unity. Thus, resetting not only shortens the mean service time but also\textit{ reduces} the relative \textit{stochastic fluctuations} around this mean, hence providing a double advantage. 

\section{M/G/1 Queues with sharp service resetting} \label{sharp resetting sec}
So far, we considered queues with resetting at a constant rate. In what follows, we consider a different extensively investigated resetting strategy, namely sharp (a.k.a deterministic or periodic) resetting \cite{PalReuveniPRL17,PalJphysA, Bhat-sharp,Discrete-res,sharp1,sharp2,sharp3}. Strong motivation to study this strategy comes from the fact that in terms of mean performance, sharp resetting  either matches, or outperforms any resetting strategy of the type considered above \eref{renewal-1-main} \cite{PalReuveniPRL17}.

To see this, let us now consider an M/G/1 queue where service is reset at fixed time intervals of length $\tau$. This simply means that the resetting time $R$ in \eref{renewal-1-main} is taken from the distribution $f_R(t)=\delta(t-\tau)$. Equations (\ref{first-mom}) and (\ref{sec-mom}), can then be simplified to give (\aref{sharp-appendix})
\bea
\langle S_\tau  \rangle &=& \frac{\int_0^\tau~dt~q_S(t)}{1-q_S(\tau)}, \label{mean sharp restart} \\
\langle S^2_\tau  \rangle &=& \Big[ 2(1-q_S(\tau))\int_0^\tau dt~ tq_S(t) \nonumber \\
&+&2\tau q_S(\tau)~\int_0^\tau dt~ q_S(t) \Big]/\left[1-q_S(\tau)\right]^2, \label{second moment sharp restart}
\eea
where $q_S(\tau)=1-\int_0^\tau~dt~f_S(t)$ is the survival function associated with the underlying service time.

\begin{figure}[t]
\includegraphics[scale=0.85]{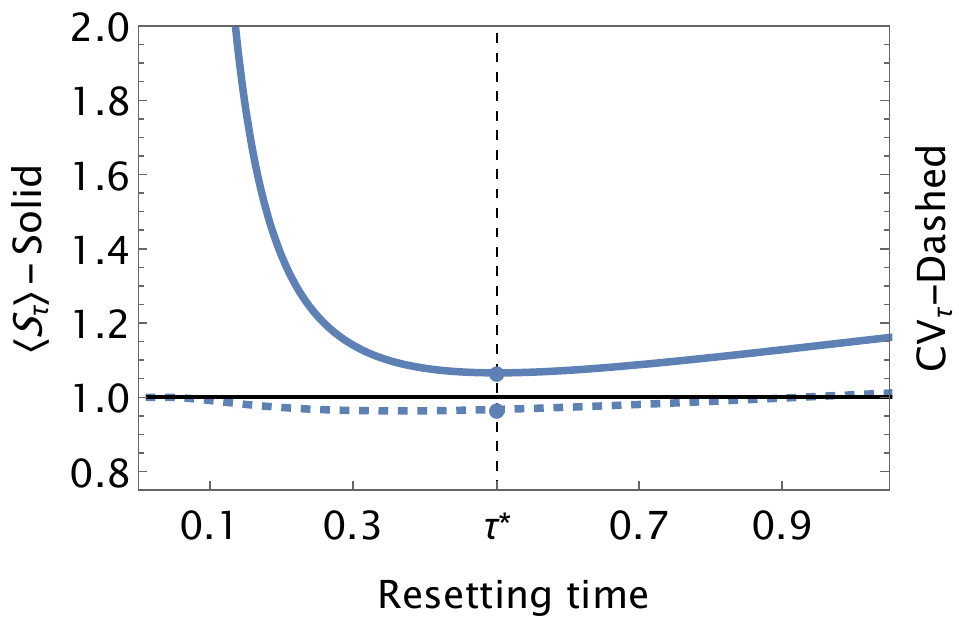}
\caption{The mean (solid line) and $CV$ (dashed line) of the service time with sharp resetting as a function of the resetting time. Plots were made, using Eqs. (\ref{mean sharp restart}) and (\ref{second moment sharp restart}), for an underlying service time taken from the Inverse Gaussian distribution whose density is given by $f_S(t) = \sqrt{\gamma / 2\pi t^3} e^{-\gamma (t-\mu)^2 / 2\mu ^2 t},~t>0$. Here, $\mu = 2.5$ and $\gamma = 0.5$. Observe that the mean service time with resetting, $\langle S_\tau \rangle$, is minimized at an optimal resetting time $\tau^* \simeq 0.501$, which was found using \eref{tau* equation}. At this optimal resetting time we have  $CV_{\tau^*}\leq 1$.}
\label{fig3}
\end{figure}

The utilization of this queue is then given by $\rho_\tau=\lambda\langle S_\tau \rangle$, and the squared coefficient of variation of the service time is  $CV_\tau^2=\text{Var}(S_\tau)/\langle S_\tau \rangle^2$. We can now obtain the mean queue length under service resetting by replacing $\rho$ by $\rho_\tau$, and $CV^2$ by $CV_\tau^2$, in \eref{PK-1}. This yields  
\bea
\langle N_{\tau} \rangle=\frac{\rho_\tau}{1-\rho_\tau}+\frac{\rho_\tau^2}{2(1-\rho_\tau)}\left( CV_\tau^2-1  \right)~.
\label{PK-N-tau}
\eea
In Fig. \ref{fig3}, we consider a representative example for the effect of sharp resetting on an M/G/1 queue with large fluctuations in service time. Starting from the same service time distribution used in Fig. \ref{fig2} ($CV>1$), we introduce resetting and plot the mean service time $\langle S_\tau \rangle$ vs. the resetting time. Once again, we observe that $\langle S_\tau \rangle$ obtains a minimum at an optimal resetting time $\tau^*$. In addition, note that at this optimal resetting time we have  $CV_{\tau^*}<1$. As for the case of Poissonian resetting, this observation is not specific to the service time distribution considered in Fig. \ref{fig3}, but is rather a universal property. Namely, it can be shown that \cite{PalReuveniPRL17}
\bea
CV_{\tau^*} \leq 1,
\label{CV-sharp-optimal}
\eea
for an optimal resetting time $\tau^*$ which brings $\langle S_{\tau^*} \rangle$ to a global minimum.
As before, $\tau^*$ can be found by setting
\bea
\frac{d \langle S_\tau  \rangle}{d\tau} \bigg|_{\tau^*}=0~, \label{Sharp resetting opt}
\eea
which by substitution of \eref{mean sharp restart} into \eref{Sharp resetting opt}, boils down to
\bea
q_S({\tau}^*)-q^2_S({\tau}^*)+q'_S({\tau}^*)\int_0^{{\tau}^*}~dt~q_S(t)=0. \label{tau* equation}
\eea

Combining \eref{CV-sharp-optimal} with the Pollaczek-Khinchin formula given in \eref{PK-N-tau}, we have
\bea
\langle N_{\tau^*} \rangle \leq ~\frac{\rho_{\tau^*}}{1-\rho_{\tau^*}}.
\label{PK-N-tau-opt}
\eea
It is interesting to compare Eqs. (\ref{PK-N-tau-opt}) and (\ref{PK-1-r*}). To this end, we recall that the mean first passage time under optimal sharp resetting is always smaller or equal than that obtained under optimal Poissonian resetting \cite{PalReuveniPRL17}. Translating this result to the language of queueing, we have $\langle S_{\tau^*} \rangle \leq \langle S_{r^*} \rangle$, which implies $\frac{\rho_{\tau^*}}{1-\rho_{\tau^*}} \leq \frac{\rho_{r^*}}{1-\rho_{r^*}}$,  by monotonicity. We thus have 
\bea
\langle N_{\tau^*} \rangle \leq \langle N_{r^*} \rangle. \label{PK-N-tau-opt-2}
\eea

Equation (\ref{PK-N-tau-opt-2}) is of great importance as it reveals that sharp resetting offers an additional improvement compared to the Poissonian resetting strategy. Once again, we see that whenever $CV>1$ for the original service time, the mean number of jobs in the queue can be reduced by resetting service at an optimal time. When doing so, one also reduces the coefficient of variation of the optimally restarted service time below unity. Thus, sharp resetting further shortens the mean service time, while also reducing the relative stochastic fluctuations around this mean compared to Poissonian resetting. We now set out to illustrate the results obtained in sections \ref{poisson resetting sec} and \ref{sharp resetting sec} on several case studies.
 
\section{Examples}
\label{examples}
To demonstrate the power of our approach, we now consider two  well-known service time distributions: log-normal \cite{Log-normal1,Log-normal2,Log-normal3} and Pareto \cite{Par1, Par2, Par3, Par4, Par5}, which are  now well documented in queuing theory literature \cite{Mor-Pareto,log-normal-callcenter1,log-normal-callcenter2, Pareto-Queue-1,Pareto-Queue-2,Pareto-Queue-3,HeavyTail1, HeavyTail2, HeavyTail3, HeavyTail4}. In what follows, we first describe the effect of restart in the case of log-normal service times and  discuss Pareto service times next.

\subsection{Log-normal service time}
\label{Log-normal service}
Consider the case of on an M/G/1 queue whose service time $S$ is log-normally distributed
\bea
f_S(t)=\frac{1}{\sqrt{2\pi} \sigma t}~e^{-\frac{(\ln t-\mu)^2}{2\sigma^2}}, \label{log normal dist}
\eea
for $t>0$, where $\mu \in (-\infty, \infty)$ and $\sigma>0$.
The survival function is then given by
\bea
q_S(t)=\text{Pr}(S>t)=\begin{cases}
 \frac{1}{2} \text{Erfc}\left(-\frac{\mu -\log (t)}{\sqrt{2} \sigma}\right) & t>0 \\
 1 & t \leq 0
\end{cases} \label{Survival LogNormal}
\eea
The mean and variance of the service time in this case are given by
\bea
&\langle S \rangle&~=~e^{\mu + \frac{\sigma^2}{2}}, \label{mean log normal} \\ 
&\text{Var}(S)&~=~\left(e^{\sigma^2} -1\right)e^{2\mu+\sigma^2}, \label{var log normal}
\eea
such that 
\bea
CV^2=e^{\sigma^2} - 1~, \label{CV log normal}
\eea
which is independent of $\mu$. In the discussion below, we set the mean service time $\langle S \rangle$ to be fixed, and vary $\sigma$, which controls the relative magnitude of fluctuations in service time via \eref{CV log normal}. Setting $\langle S \rangle$ and $\sigma$, $\mu$ is uniquely  determined by \eref{mean log normal}.

\begin{figure}[t]
\includegraphics[scale=0.6]{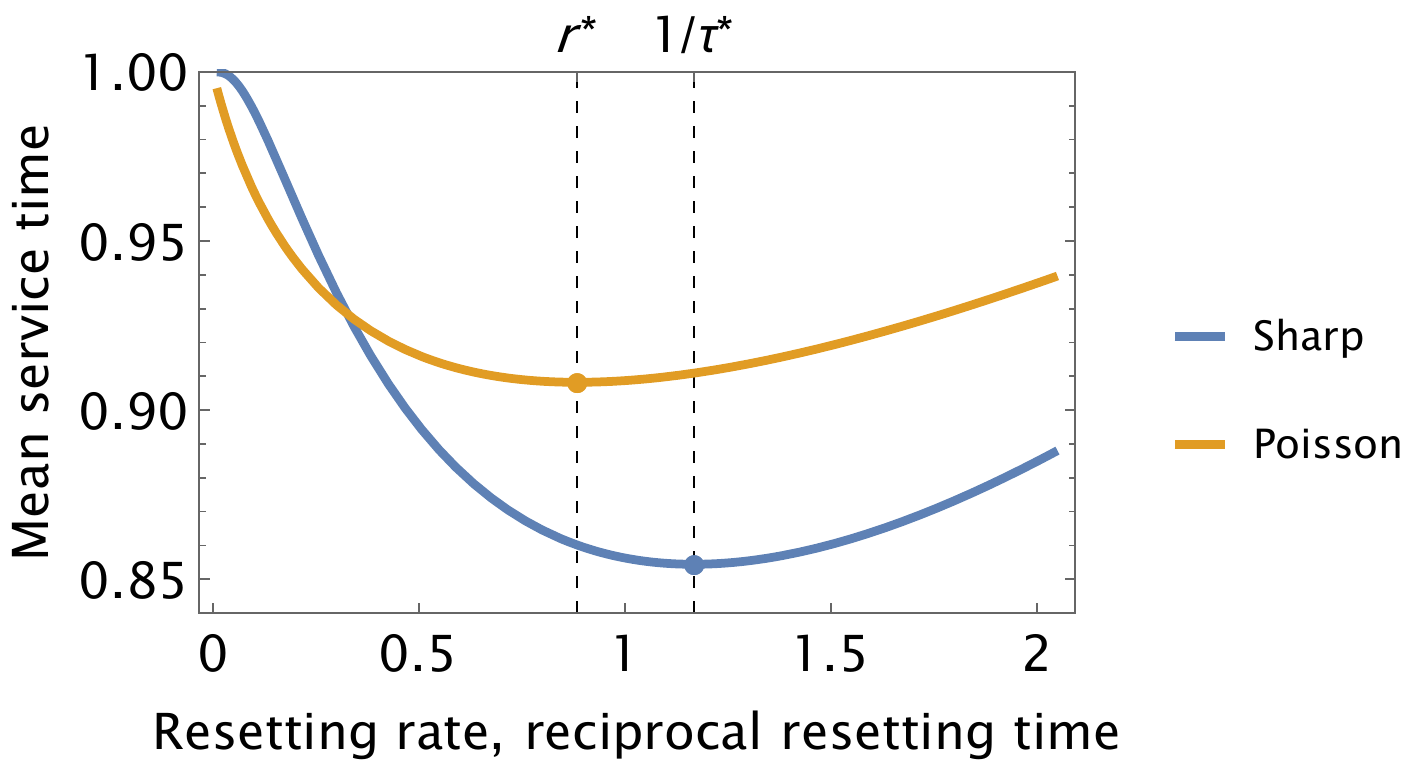}
\caption{The mean service time with Poissonian (orange) and sharp (blue) resetting, as a function of the resetting rate and reciprocal resetting time. Plots were made, using Eqs. (\ref{mean under restart}) and (\ref{mean sharp restart}), for an underlying service time taken from the log-normal distribution whose density is given by \eref{log normal dist}. Here, $\langle S \rangle = 1$ and $\sigma = 1.05$, yielding $\mu = -0.28125$. The optimal resetting rate, $r^*$, and resetting time, $\tau^*$, are indicated.} 
\label{fig4}
\end{figure}

For Poissonian resetting, the mean service time is given by Eq. (\ref{mean under restart}), which requires the Laplace transform of the log-normal distribution. The latter, does not have an analytical closed-form, but it can be evaluated numerically for any choice of parameters. In the case of sharp resetting, the mean service time is given by \eref{mean sharp restart}, which requires the survival function $q_S(t)$ given in \eref{Survival LogNormal}. Here too, the mean service under resetting can be computed by numerical  evaluation of the required integrals. In Fig. \ref{fig4}, we set $\langle S \rangle = 1$ and $\sigma = 1.05$, and plot the mean service time under Poissonian and sharp resetting. In both cases, a minimum is obtained at an optimal resetting rate or time, depending on the resetting scheme. Observe that the optimal mean service time under sharp resetting is indeed lower than that obtained for Poissonian resetting. 

To find the optimal resetting rate in Fig. \ref{fig4}, we solve  
\bea 
\mathcal{H}(\langle S \rangle,\sigma,r^*) = 0, \label{trans r*}
\eea 
where $\mathcal{H}(\langle S \rangle,\sigma,r^*)$ denotes the left hand side of \eref{r* equation}. This gives $r^* \simeq 0.885$ for the given parameters. A similar minimization procedure can be carried out for sharp resetting. Substitution of the survival function for the log-normal distribution given in \eref{Survival LogNormal} into \eref{tau* equation} yields the following equation for the optimal resetting time $\tau^*$
\begin{align}
    \mathcal{F}(\langle S \rangle,\sigma,\tau^*)=0,
    \label{transc-1}
\end{align}
where $\mathcal{F}(\langle S \rangle,\sigma,\tau^*)$ denotes the left hand side of \eref{tau* equation}. This gives $\tau^*\simeq 0.857$.

\begin{figure*}[t]
\includegraphics[width=9.5cm,height=5cm]{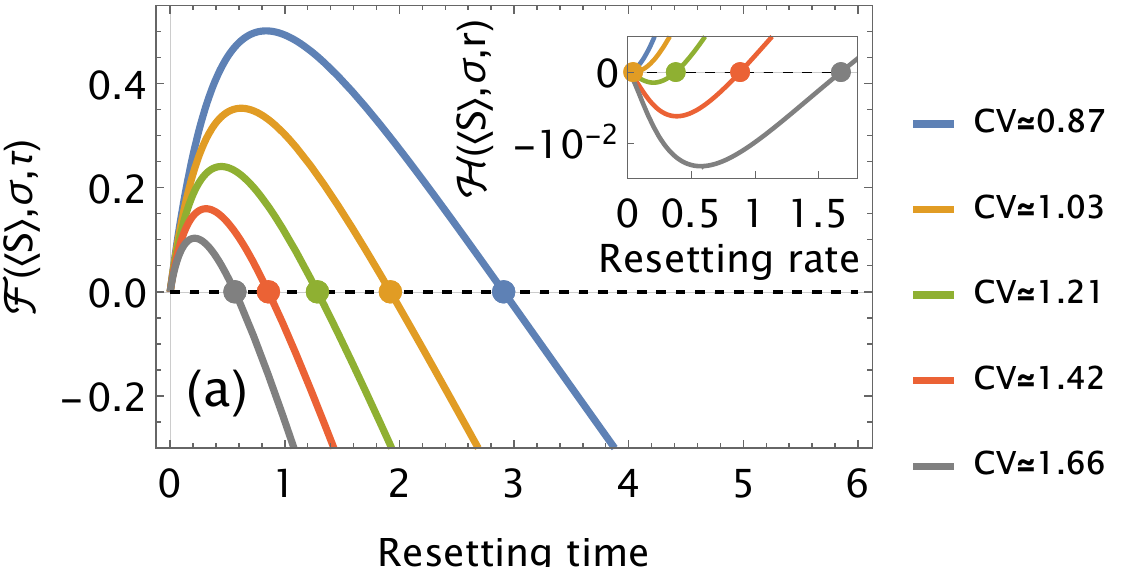}
\includegraphics[width=7.7cm,height=5cm]{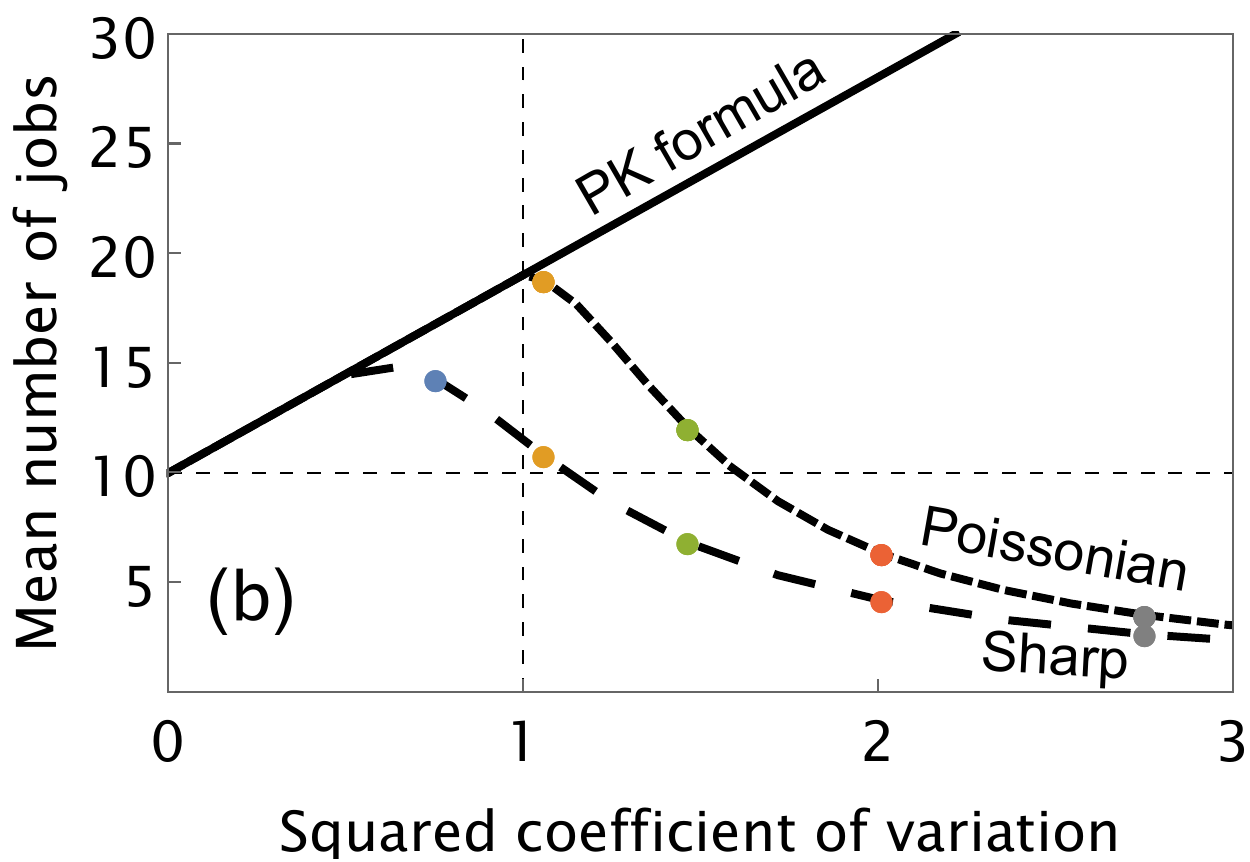}
\caption{Panel (a): Solutions of Eqs. (\ref{trans r*}) and (\ref{transc-1}) for the Log-normal service time distribution under Poissonian (inset) and sharp resetting (main). Here, we fix $\langle S \rangle=1$ for the mean service time without resetting and vary $\sigma$ to control relative stochastic fluctuations, $CV$, via \eref{CV log normal}. The intersection points of the curves with the dashed horizontal line going through the origin give the optimal restart times $\tau^*$ (optimal restart rates $r^*$ in the inset). Note that higher values of $\sigma$ yield lower optimal resetting times (higher resetting rates). Panel (b): The mean number of jobs in the queue as a function of the underlying $CV^2$ of the service time distribution. The Pollaczek-Khinchin formula gives the familiar linear dependence of \eref{PK-1}. Also plotted are the behaviours for the optimal Poissonian and sharp resetting protocols, with colored circles matching their counterparts in panel (a). Strong deviations from the Pollaczek-Khinchin behaviour of the non-restarted case are observed. While resetting provides no advantage for low $CV$ values, it can drastically reduce queue lengths when $CV$ is high.}
\label{log-roots}
\end{figure*}

Once in hand, the optimal resetting rate $r^*$ can be substituted into Eqs. (\ref{mean under restart}) and (\ref{secmom under restart}), to obtain the mean and variance of the service time under optimal Poissonian resetting. Similarly, $\tau^*$ can be substituted into Eqs. (\ref{mean sharp restart}) and (\ref{second moment sharp restart}), for the mean and variance of the service time under optimal sharp resetting. Once these quantities are known, and given the arrival rate $\lambda$, the Pollaczek-Khinchin formula can be used directly to obtain the the mean number of jobs in a queue with optimal service resetting. 

In Fig. \ref{log-roots}, we extend the analysis presented in Fig. \ref{fig4}. Setting $\langle S \rangle = 1$, we compute the optimal resetting rate and time for various values of the parameter $\sigma$ (Fig. \ref{log-roots}a), which controls the relative fluctuations in the log-normal service time via \eref{CV log normal}. We then compare between the mean number of jobs in a queue without service resetting to that obtained when  optimal Poissonian and sharp resetting are applied (\fref{log-roots}b). We observe that resetting can drastically shorten queues when stochastic fluctuations in the underlying service time are large (high $CV$). To this end, note that optimal sharp resetting outperforms optimal Poissonian resetting as predicted by \eref{PK-N-tau-opt-2}. Also, note that for high $CV$ values the mean queue lengths at optimal resetting drop below the mean queue length in the absence of service time fluctuations ($CV=0$). This remarkable feature illustrates how resetting turns the problem of service time fluctuations into an advantageous benefit.

\subsection{Pareto service time}
\label{Pareto service}
We now consider the case of on an M/G/1 queue whose service time $S$ is Pareto distributed
\bea
f_S(t)=\frac{\alpha L^{\alpha}}{t^{\alpha+1}}, \label{Pareto dist}
\eea
for $t\geq L$, where $\alpha,L>0$.
The survival function is given by
\bea
q_S(t)~=~\begin{cases}
 \left(\frac{t}{L}\right)^{-\alpha } & t\geq L \\
 1 & t<L
\end{cases} \label{Survival Pareto}.
\eea

The mean and variance of the service time in this case are given by
\bea
&\langle S \rangle&~=~\begin{cases}
\infty & \text{for}\;\alpha\leq1\\
\frac{\alpha L}{\alpha-1} & \text{for} \;\alpha>1
\end{cases}, \label{Mean Pareto}\\ 
&\text{Var}(S)&~=~\begin{cases}
\infty & \text{for} \;\alpha\leq2\\
\frac{\alpha L^{2}}{(\alpha-1)^{2}(\alpha-2)} & \text{for} \;\alpha>2
\end{cases}, \label{Variance Pareto}
\eea
such that
\bea
CV^2~=~\frac{1}{\alpha (\alpha-2)}, \label{CV pareto}
\eea 
which is independent of $L$. Similar to the previous example, we now set the mean service time $\langle S \rangle$ to be fixed, and vary $\alpha$, which controls the relative magnitude of fluctuations in service time via \eref{CV pareto}. Setting $\langle S \rangle$ and $\alpha$, $L$ is uniquely  determined by \eref{Mean Pareto}.

For Poissonian resetting, the mean service time is given by Eq. (\ref{mean under restart}), which requires the Laplace transform of the Pareto distribution. This is given by
\bea
\tilde{S}(r)=\alpha (Lr)^{\alpha} \Gamma(-\alpha,Lr), \label{Laplace Pareto}
\eea 
where $\Gamma(a,x)$ is the upper incomplete gamma function
\bea 
\Gamma(a,x) = \int_x^\infty~dt~ t^{\alpha - 1}e^{-t}.
\eea 
By substituting the above into Eqs. (\ref{mean under restart}) and (\ref{secmom under restart}), we get the following expressions for the mean
\bea
\langle S_r \rangle &=& \frac{1-\alpha (Lr)^{\alpha} \Gamma(-\alpha,Lr)}{\alpha r (Lr)^{\alpha} \Gamma(-\alpha,Lr)}, \label{mean Sr Pareto}
\eea
and second moment
\bea 
\langle S_r^2 \rangle = \frac{2\alpha (\alpha -1)
   (L r)^{\alpha } \Gamma (-\alpha ,L r)-2\alpha e^{-L r} +2 }{(\alpha r) ^2(L r)^{2 \alpha }\Gamma (-\alpha ,L r)^2},
\eea 
of the service time under Poissonian resetting.

In the case of sharp resetting, the first two moments of the service time are given by Eqs. (\ref{mean sharp restart}) and (\ref{second moment sharp restart}). Substituting the survival function (\ref{Survival Pareto}) into Eqs. (\ref{mean sharp restart}) and (\ref{second moment sharp restart}) yields 
\bea
\langle S_\tau  \rangle &=& \frac{L\alpha}{\alpha-1} + \frac{L\alpha -\tau}{(\alpha-1)\left[\left(\frac{\tau}{L}\right)^{\alpha}-1\right]}, \label{mean Pareto sharp restart} \\
\langle S^2_\tau  \rangle &=& \frac{2 \tau ^2 L^{\alpha } \left(L^{\alpha }-(\alpha -1) \tau ^{\alpha }\right)}{(\alpha -2) (\alpha -1) \left(\tau ^{\alpha }-L^{\alpha }\right)^2}+\frac{2 \alpha  \tau ^{\alpha +1} L^{\alpha +1}}{(\alpha -1) \left(\tau ^{\alpha }-L^{\alpha}\right)^2} \nn \\
&~&+\frac{\alpha  L^2 \tau ^{\alpha }}{(\alpha -2) \left(\tau ^{\alpha }-L^{\alpha }\right)} .\label{second moment Pareto sharp restart}
\eea
Note that in the above we only considered resetting times $\tau > L$ as the support of the Pareto distribution is given by $t\geq L$. Also, note that \eref{mean Pareto sharp restart} is valid for $\alpha \neq 1$. Similarly, in \eref{second moment Pareto sharp restart} we require $\alpha \neq 1,2$. These singular cases require separate treatment, following similar footsteps.

\begin{figure}[t]
\includegraphics[scale=0.6]{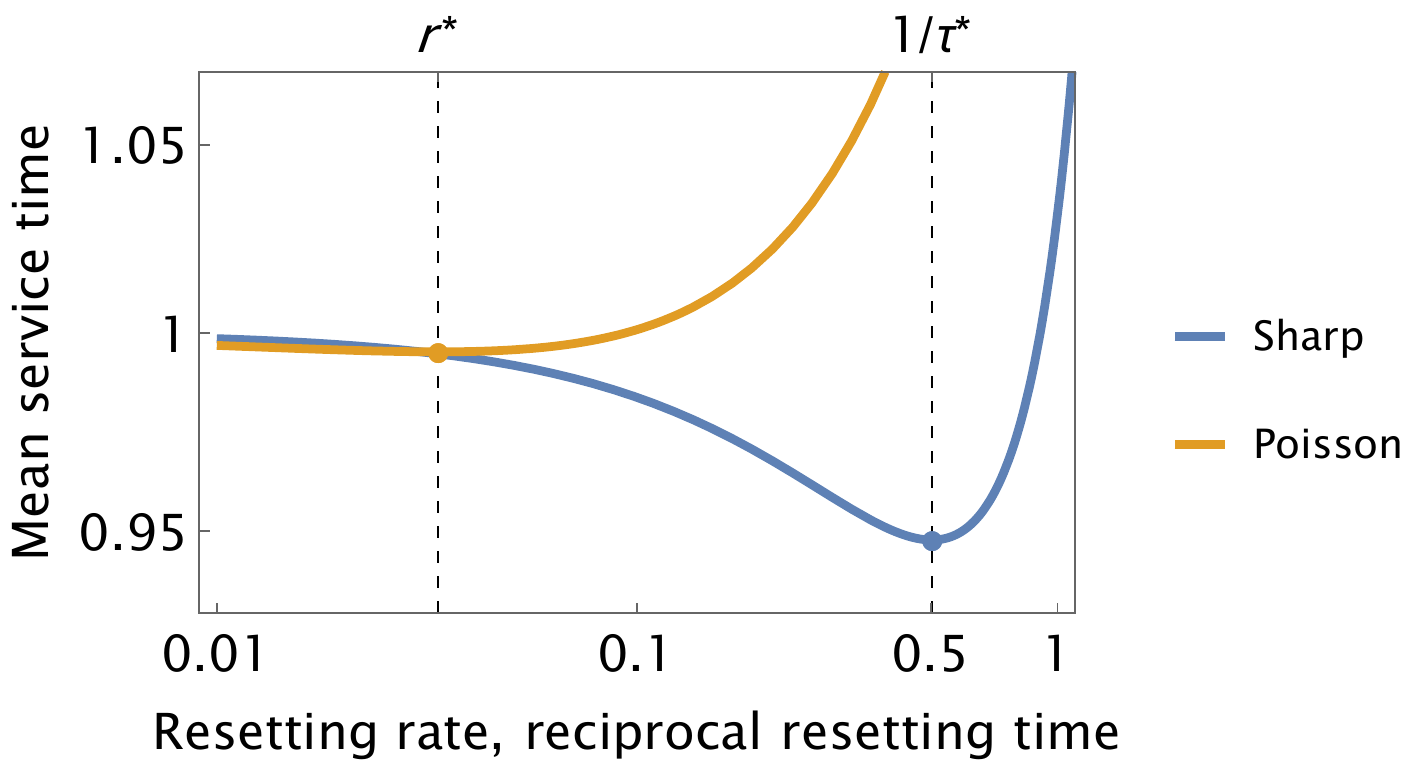}
\caption{The mean service time with Poissonian (orange) and sharp (blue) resetting, as a function of the resetting rate and reciprocal resetting time. Plots were made, using Eqs. (\ref{mean Sr Pareto}) and (\ref{mean Pareto sharp restart}), for an underlying service time taken from the Pareto distribution whose density is given by \eref{Pareto dist}. Here, $\langle S \rangle = 1$ and $\alpha = 2.1$, yielding $L \simeq 0.524$. The optimal resetting rate, $r^*$, and resetting time, $\tau^*$, are indicated.}
\label{fig5}
\end{figure}

In Fig. \ref{fig5}, we set $\langle S \rangle = 1$ and $\alpha = 2.1$, and plot the mean service time under Poissonian and sharp resetting. In both cases, a minimum is obtained at an optimal resetting rate or time, depending on the resetting scheme. Once again, the optimal mean service time under sharp resetting is indeed lower than that obtained for Poissonian resetting. 

To find the optimal resetting rate in Fig. \ref{fig5}, we solve  
\bea 
\mathcal{S}(\langle S \rangle,\alpha,r^*) = 0, \label{Pareto trans r*}
\eea 
where $\mathcal{S}(\langle S \rangle,\alpha,r^*)$ denotes the left hand side of \eref{r* equation}, after substituting the Laplace transform of \eref{Laplace Pareto}. This gives $r^* \simeq 0.034$. A similar minimization procedure can be carried out for sharp resetting. Substitution of the survival function for the Pareto distribution given in \eref{Survival Pareto} into \eref{tau* equation} yields the following equation for the optimal resetting time $\tau^*$
\bea
\mathcal{G}(\langle S \rangle,\alpha,\tau^*)=
{\tau}^*,
\label{transc-2}
\eea
where 
\bea
\mathcal{G}(\langle S \rangle,\alpha,\tau)=\left(\frac{{\tau} }{L}\right)^{\alpha } \left(-\alpha  {\tau} +\alpha ^2 L+{\tau} \right).
\eea
Solving gives $\tau^* \simeq 1.990$. 

The optimal resetting rate $r^*$ can be substituted into Eqs. (\ref{mean under restart}) and (\ref{secmom under restart}), to obtain the mean and variance of the service time under optimal Poissonian resetting. Similarly, $\tau^*$ can be substituted into Eqs. (\ref{mean sharp restart}) and (\ref{second moment sharp restart}), for the mean and variance of the service time under optimal sharp resetting. Once these quantities are known along with the arrival rate $\lambda$, the Pollaczek-Khinchin formula can be used directly to obtain the the mean number of jobs in a queue with optimal service resetting.

\begin{figure*}[t]
\includegraphics[width=9cm,height=5cm]{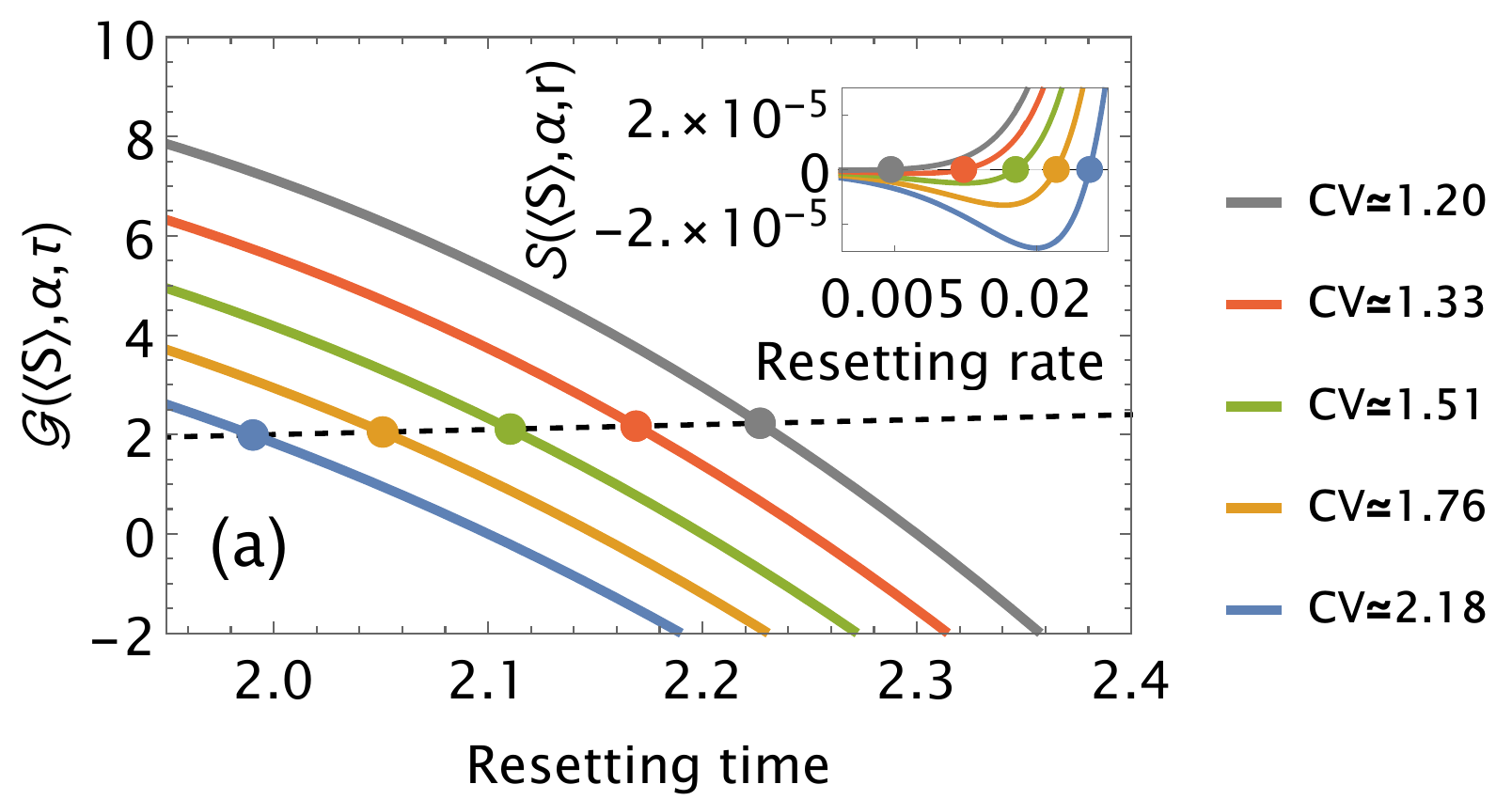}
\includegraphics[width=7.7cm,height=5cm]{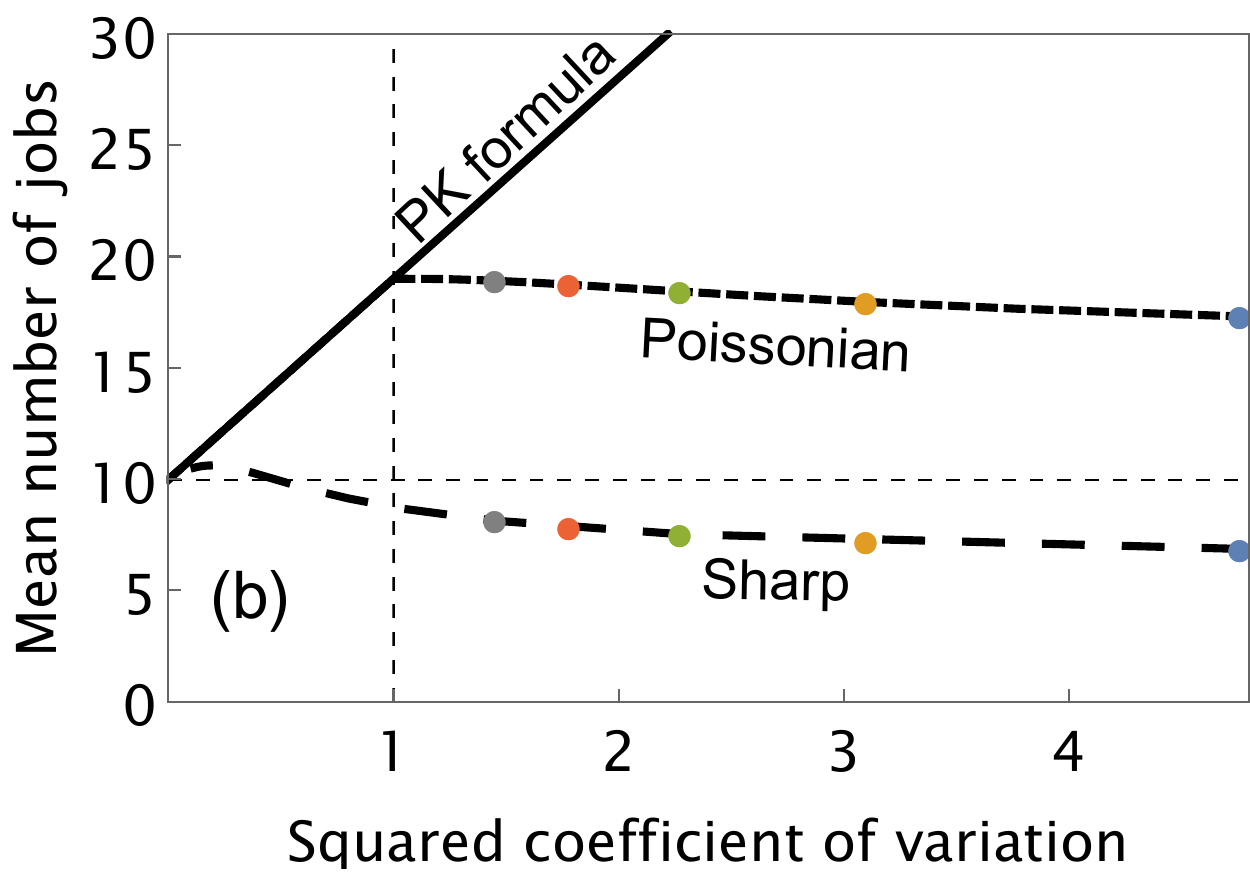}
\caption{Panel (a): Solutions of Eqs. (\ref{Pareto trans r*}) and (\ref{transc-2}) for the Pareto service time distribution under Poissonian (inset) and sharp resetting (main). The intersection points of the curves for different $\alpha$ (i.e., for different $CV$) with the dashed horizontal line through the origin give the optimal restart time $\tau^*$ (optimal restart rate $r^*$ in the inset). Panel (b): The mean number of jobs in the queue as a function of the underlying $CV^2$ of the service time distribution. The Pollaczek-Khinchin formula gives the familiar linear dependence of \eref{PK-1}. Also plotted are the behaviours for the optimal Poissonian and sharp resetting protocols, with colored circles matching their counterparts in panel (a). Strong deviations from the Pollaczek-Khinchin behaviour of the non-restarted case are observed. While resetting provides no advantage for low $CV$ values, it can drastically reduce queue lengths when $CV$ is high. }
\label{pareto-roots}
\end{figure*}

In Fig. \ref{pareto-roots}, we extend the analysis presented in Fig. \ref{fig5}. Setting $\langle S \rangle = 1$, we compute the optimal resetting rate and time for various values of the parameter $\alpha$ (Fig. \ref{pareto-roots}a), which controls the relative fluctuations in the Pareto service time via \eref{CV pareto}. We then compare between the mean number of jobs in a queue without service resetting to that obtained when  optimal Poissonian and sharp resetting are applied (\fref{pareto-roots}b). Like in the previous example, here too, we observe that resetting can induce shorter queues when the underlying service time variability ($CV$) is large. Noteworthy in this case is the dominance of sharp resetting over Poissonian resetting as predicted by \eref{PK-N-tau-opt-2}. Moreover,  for high $CV$ values the mean queue length at optimal resetting drops below the mean queue length in the absence of service time fluctuations ($CV=0$). This hallmark property shows how resetting can take advantage of large stochastic fluctuations in service time and expedite the process completion.

\section{Conclusions and outlook}\label{Conc}
\noindent Regulating the number of jobs in a queue is an integral part of performance modeling and optimization of queuing systems. One problem that arises in this context is that large stochastic fluctuations in service times lead to significant backlogs and delays. In this paper, we showed how this problem can be mitigated by service resetting. Indeed, we have shown that when applied to servers with intrinsically highly service time variability, resetting can dramatically reduce queue lengths and job waiting times.

Our analysis was based on the canonical M/G/1 queuing model in which jobs arrive to the queue following a Poisson process and service times come from a general distribution. The renowned Pollaczek-Khinchin formula (Eq. \ref{PK-1}) asserts that the mean number of jobs in this queue grows linearly with the squared coefficient of variation, $CV^2$, of the service time. Employing the recently developed framework of first-passage under restart \cite{ReuveniPRL16,PalReuveniPRL17} to the M/G/1 queue, we showed that Poissonian service resetting reduces both the mean and variance of the overall service time when $CV>1$, i.e., exactly when service time fluctuations start to become a major source of concern. Sharp service resetting, a.k.a deterministic or periodic, performs even better and could in some cases lower the mean and variance of the service time even when $CV<1$. In both cases, the net result is much shorter queues and examples were given to show that mean queue lengths can drop well below those attained for servers with no  service time fluctuations (deterministic service time, $CV=0$). Service resetting can thus turn a well-known drawback of queueing systems into a favourable advantage. 

Our work is the first step towards the application of resetting as a fluctuations mitigation strategy in queueing systems. Some future research directions are discussed below. The analysis presented above yielded an analytical result for the mean number of jobs in an M/G/1 queue under service resetting, thus generalizing the Pollaczek-Khinchin formula to this case. In the future, it would be interesting to generalize this result beyond the mean, so as to give the full distribution of the queue length under service resetting. Another interesting direction would be to generalize the analysis to other queueing systems, e.g., the M/G/k queue which has $k$-servers or to other multi-server systems. 

Going to queues with non-Markovian arrivals is also of interest, as realistic job arrival processes need not be Poissonian. To this end, we recall Kingman's approximation formula \cite{Kingman}, which asserts that the G/G/1 queue suffers from the same linear dependence of the mean number of jobs on the squared coefficient of variation of the service time. Thus, applying service resetting to this queue will qualitatively have the same effect as in the simpler M/G/1 queue. Moreover, as the arrival process is decoupled from the service process, the results derived in Secs. \ref{Sec 3}-\ref{sharp resetting sec} above can be applied directly to the G/G/1 queue, with the only difference being Kingman's approximation formula substituting for  Pollaczek-Khinchin. Indeed, this can be done as both formulas only depend on the first two moments of the service time distribution. Conclusions coming from this work thus apply broadly to single-server queuing systems, and their ramifications on multi-server queues and queue networks remain to be worked out and understood in detail.    

\begin{acknowledgments}
Arnab Pal is indebted to Tel Aviv University (via the Raymond and Beverly Sackler postdoc fellowship and the fellowship from Center for the Physics and Chemistry of Living Systems) where the project started. Shlomi Reuveni acknowledges support from the Israel Science Foundation (grant No. 394/19). This project has received funding from the European Research Council (ERC) under the European Union’s Horizon 2020 research and innovation programme (Grant agreement No. 947731).
\end{acknowledgments}

\appendix

\section{Service time under restart}
\label{SR-PDF-LT}
\noindent 
In this section, we briefly review a few essential results that were derived employing the framework of first passage under restart in \cite{PalReuveniPRL17}. We start by recalling that the distribution of service time under restart in Laplace space reads
\cite{PalReuveniPRL17}
\bea
\tilde{S}_R(s)=\frac{\text{Pr}(S<R)\tilde{S}_{\text{min}}(s)}{1- \text{Pr}(R \leq S) \tilde{R}_{\text{min}}(s)}~,
\label{LT-genericR}
\eea
where $\tilde{S}_R(s)=\int_0^\infty~dt~e^{-st}~f_{S_R}(t)$. Moreover, we have defined two auxiliary random variables
\bea
R_{\text{min}} &\equiv& \{R|R\leq S \}~,\nonumber\\
S_{\text{min}} &\equiv& \{S|S<R \}~.
\label{rmin-tmin-defn}
\eea
In words, $R_{\text{min}}$ is the restart time conditioned on the event that restart occurs before the service is over. Similarly, $S_{\text{min}}$ is the time conditioned that service occurred prior to a restart. The probability density functions of $R_{\text{min}}$ and $S_{\text{min}}$ are given by \cite{PalReuveniPRL17}
\begin{align}
f_{R_{\text{min}}}(t) &= \frac{f_R(t) \int_t^\infty~dt'~f_S(t')}{\text{Pr}(R\leq S)}=\frac{f_R(t) \text{Pr}(S>t)}{\text{Pr}(R\leq S)}~, \\
f_{S_{\text{min}}}(t) &= \frac{f_S(t) \int_t^\infty~dt'~f_R(t')}{\text{Pr}(S<R)}=\frac{f_S(t) \text{Pr}(R>t)}{\text{Pr}(S<R)}~.
\label{rmin-tmin-PDF}
\end{align}
The moments can now easily be computed from \eref{LT-genericR} by noting that 
\bea
\langle S_R^n \rangle=(-1)^n \frac{d^n}{ds^n} \tilde{S}_R(s)|_{s \to 0} ,
\eea
which gives \cite{PalReuveniPRL17}
\begin{align}
\langle S_R  \rangle &= \frac{\langle \text{min}(S,R) \rangle}{\text{Pr}(S<R)} \label{mean SR}\\
\langle S_R^2  \rangle &= \frac{\langle \text{min}(S,R)^2 \rangle}{\text{Pr}(S<R)}+
\frac{2 \text{Pr}(R \leq S) \langle   R_{\text{min}}\rangle \langle \text{min}(S,R) \rangle}{\text{Pr}(S<R)^2}, \label{second moment Sr}
\end{align}
where $\langle   R_{\text{min}}\rangle$ can be computed directly from \eref{rmin-tmin-PDF}. The other components can also be systematically derived with the knowledge of individual time densities.
For example,
the numerator $\langle \text{min}(S,R) \rangle$ in \eref{mean SR} is given by
\bea
\text{Pr}(\text{min}(S,R)\leq t) = 1-\text{Pr}(S>t)\text{Pr}(R>t).
\eea
and the denominator in \eref{mean SR} is given by
\bea
\text{Pr}(S<R)&=&\int_0^\infty~dt~f_R(t)\text{Pr}(S<t) \nonumber \\
&=&\int_0^\infty~dt~f_R(t)
\int_0^t~dt'~f_S(t'). \label{Pr S R}
\eea

\subsection{Moments of service time under Poissonian resetting} \label{A1}

Consider the case of Poissonian resetting. When the restart time $R$ is Poissonian, its probability density function is given by 
\bea
f_R(t)= r~e^{-rt}, \label{Poissonian density}
\eea
and the cumulative distribution function is given by
\bea
\text{Pr}(R \leq t)= 1-e^{-rt}, \label{Cumulative density}
\eea
where $r$ is the resetting rate.

In the case of Poissonian resetting, the cumulative distribution function of $\text{min}(S,R)$ can be written as followed
\bea
\text{Pr}(\text{min}(S,R)\leq t)= 
1-\text{Pr}(S>t)~e^{-rt}. \label{min cumulative Poissonian}
\eea
Using the following formula for the expectation value of non-negative random variable 
\bea
\langle X \rangle = \int_0^{\infty} dt~ q_X(t), \label{integral formula}
\eea
where $q_X(t)=\text{Pr}(X>t)$ is the survival function of $X$, one can easily show that
\bea
\langle \text{min}(S,R) \rangle &=& \int_0^{\infty} dt~ \text{Pr}(S>t)\text{Pr}(R>t) \nn \\ &=&\int_0^{\infty} dt~ \text{Pr}(S>t)e^{-rt} \nn \\
&=&\int_0^{\infty} dt~ \left[ 1 - \text{Pr}(S<t) \right] e^{-rt} \nn \\
&=&\int_0^{\infty} dt~e^{-rt} - \int_0^{\infty} dt~ \text{Pr}(S<t)e^{-rt} \nn \\
&=& \frac{1}{r} - \int_0^{\infty} dt~ e^{-rt} \left( \int_0^{t} dt'~ f_S (t') \right) \nn \\
&=& \frac{1}{r} - \frac{\tilde{S}(r)}{r} = \frac{1-\tilde{S}(r)}{r}, \label{mean min Poiss}
\eea 
where in the second to last step we used a known formula for the Laplace transform of a time-domain integration: $\int_0^{\infty} dt~ \left (\int_0^{t}d\tau~g(\tau) \right ) e^{-rt} = \frac{\tilde{g}(r)}{r}$, where $\tilde{g}(r)$ is the Laplace transform of $g(t)$. Similarly, we can use \eref{Pr S R} to find 
\bea 
\text{Pr}(S<R)&=&\int_0^\infty~dt~r~e^{-rt}\text{Pr}(S<t) \nn \\
&=& r \int_0^\infty~dt~e^{-rt}~\text{Pr}(S<t) \nn \\
&=& \tilde{S}(r), \label{Prob SR Poiss}
\eea 

Substituting Eqs.(\ref{mean min Poiss}) and (\ref{Prob SR Poiss}) into Eq.(\ref{mean SR}), we get the following formula for the mean service time under Poissonian resetting  
\begin{equation}
\langle S_r \rangle = \frac{1-\tilde{S}(r)}{r \tilde{S}(r)},
\label{Mean Poiss}
\end{equation}
which was used in the main text (see \eref{mean under restart}). This equation was previously derived in \cite{PalReuveniPRL17}.

We now turn to the derivation of formula for the second moment of the service time under Poissonian resetting, $\langle S_r^2 \rangle$. To do so, we start by deriving $\langle \text{min}(S,R)^2 \rangle$. By taking the derivative of Eq. (\ref{min cumulative Poissonian}), one can easily obtain the probability density function of $\text{min}(S,R)$
\bea
f_{\text{min}(S,R)} (t)&=& f_S(t)~e^{-rt} +r~\text{Pr}(S>t)~e^{-rt}. \label{min density Poiss}
\eea
Now, we can use the probability density function given in Eq.(\ref{min density Poiss}) to calculate $\langle \text{min}(S,R)^2 \rangle$
\begin{align}
\langle \text{min}(S,R)^2 \rangle&=\int_0^{\infty} dt~ t^2 \left( f_S(t)~e^{-rt} +r~\text{Pr}(S>t)~e^{-rt} \right) \nn \\ 
&= \int_0^{\infty} dt~ t^2 f_S(t)~e^{-rt} \nn\\
&+ r\int_0^{\infty} dt~ t^2~\text{Pr}(S>t)~e^{-rt} \nn \\ 
&= \frac{d^2\tilde{S}(r)}{dr^2} + r\int_0^{\infty} dt~ t^2 \left( 1-\text{Pr}(S<t)\right)e^{-rt} \nn \\
&= \frac{d^2\tilde{S}(r)}{dr^2} +\frac{2}{r^2} - r\int_0^{\infty} dt~ t^2\text{Pr}(S<t)e^{-rt} \nn \\
&= \frac{d^2\tilde{S}(r)}{dr^2} +\frac{2}{r^2} - r \frac{d^2 \left (\frac{ \tilde{S}(r)}{r}\right )}{dr^2} \nn \\
&= \frac{2r~\frac{d\tilde{S}(r)}{dr} -2\tilde{S}(r)+2}{r^2}, \label{second mom min SR Poiss}
\end{align}
where during the derivation we used a known property of Laplace transforms: $\int_0^{\infty} dt~t^n g(t)e^{-rt} = (-1)^n \frac{d^n \tilde{g}(r)}{dr^n}$, where $n$ is a positive integer with $\tilde{g}(r)$ being the Laplace transform of $g(t)$.

We now turn to calculate $\langle R_\text{min} \rangle$. This can be explicitly done using Eq.(\ref{rmin-tmin-PDF})
\bea 
\langle R_\text{min} \rangle&=&\int_0^{\infty} dt~ tf_{R_{min}}(t) \nn \\ 
&=& \frac{1}{\text{Pr}(R\leq S)}\int_0^{\infty} dt~t~re^{-rt} \text{Pr}(S>t) \nn \\ 
&=& \frac{r}{1-\text{Pr}(S<R)}\int_0^{\infty} dt~t~e^{-rt} \left(1-\text{Pr}(S<t)\right) \nn \\ 
&=& \frac{r\left(\int_0^{\infty} dt~t~e^{-rt}-\int_0^{\infty} dt~t~e^{-rt} \text{Pr}(S<t)\right)}{1-\tilde{S}(r)} \nn \\ 
&=& \frac{r\left(\frac{1}{r^2} + \frac{d\left(\frac{\tilde{S}(r)}{r}\right)}{dr}\right)}{1-\tilde{S}(r)} \nn \\
&=& \frac{r~\frac{d\tilde{S}(r)}{dr} -\tilde{S}(r)+1}{r\left(1-\tilde{S}(r) \right)}. \label{Rmin Poiss}
\eea 

Having all the terms on the right hand side of Eq. (\ref{second moment Sr}) in hand, we can now calculate $\langle S_r^2  \rangle$
\bea
\langle S_r^2  \rangle&=&\frac{\langle \text{min}(S,R)^2 \rangle}{\text{Pr}(S<R)}+
\frac{2 \text{Pr}(R \leq S) \langle   R_{\text{min}}\rangle \langle \text{min}(S,R) \rangle}{\text{Pr}(S<R)^2} \nn \\
&=& \frac{2r~\frac{d\tilde{S}(r)}{dr} -2\tilde{S}(r)+2}{r^2\tilde{S}(r)} \nn \\
&&+ \frac{2\left(1-\tilde{S}(r) \right)\frac{r~\frac{d\tilde{S}(r)}{dr} -\tilde{S}(r)+1}{r\left(1-\tilde{S}(r) \right)}\frac{1-\tilde{S}(r)}{r}}{\tilde{S}(r)^2} \nn \\
&=& \frac{2\left( r\frac{d\tilde{S}(r)}{dr} - \tilde{S}(r) +1\right)}{r^2\tilde{S}(r)^2},
\eea
which was used in the main text (see \eref{secmom under restart}).

\subsection{Moments of service time under sharp resetting}
\label{sharp-appendix}
Consider now the case of deterministic resetting. When the restart time $R$ is deterministic, its probability density function is given by 
\bea
f_R(t)=\delta(t-\tau), \label{sharp density}
\eea
where $\tau$ is the resetting time and $\delta(t)$ is the delta function.
In the case of sharp resetting, the cumulative distribution function of $\text{min}(S,R)$ can be written as followed
\bea
\text{Pr}(\text{min}(S,\tau)\leq t)= 
1-\text{Pr}(S>t)\theta(\tau-t), \label{min cumulative}
\eea
where we have used $\text{Pr}(R > t)=\int_t^{\infty} dt'~\delta(t'-\tau) = \theta(\tau-t)$,  where $\theta$ is the Heaviside step function. Using \eref{integral formula} once again one can easily show that
\begin{align}
\langle \text{min}(S,\tau) \rangle = \int_0^{\infty} dt~ \text{Pr}(S>t)\theta(\tau-t)=\int_0^{\tau} dt~ q_S(t), \label{mean min}    
\end{align}
where $q_S(\tau)=1-\int_0^\tau~dt~f_S(t)$ is the survival function associated with the underlying service time. To derive $\text{Pr}(S<R)$ we once again use \eref{Pr S R} to find
\bea
\text{Pr}(S<R) &=& 
 \int_0^\infty~dt~\delta(t-\tau)\text{Pr}(S<t) \nn \\
&=& \int_0^\infty~dt~\delta(t-\tau)(1-q_S(t)) \nn \\
&=& 1-q_S(\tau). \label{Sharp denominator}
\eea 

Substituting Eqs.(\ref{mean min}) and (\ref{Sharp denominator}) into Eq.(\ref{mean SR}), we get the following formula for the mean service time under sharp restart  
\begin{equation}
\langle S_{\tau} \rangle= \frac{\int_0^{\tau} dt~ q_S(t)}{1-q_S(\tau)}~,
\label{T-sharp}
\end{equation}
which was used in the main text (see \eref{mean sharp restart}). This equation was previously derived in \cite{PalJphysA,sharp1}.
A discrete analog for Eq. (\ref{T-sharp}) was derived in \cite{Discrete-res}. 

We now turn to the derivation of formula for the second moment of the service time under sharp restart, $\langle S_\tau^2 \rangle$. To do so, we start by deriving $\langle \text{min}(S,\tau)^2 \rangle$. By taking the derivative of Eq. (\ref{min cumulative}), one can easily obtain the probability density function of $\text{min}(S,\tau)$
\begin{align}
f_{\text{min}(S,\tau)}(t)=-\frac{\partial q_S(t)}{\partial t}\theta(\tau-t) + q_S(t)\delta(\tau-t). \label{min density}
\end{align}
Now, we can use Eq.(\ref{min density}) to compute $\langle \text{min}(S,\tau)^2 \rangle$ which reads
\begin{align}
\langle \text{min}(S,\tau)^2 \rangle&=\int_0^{\infty} dt~ t^2 \left[-\frac{\partial q_S(t)}{\partial t}\theta(\tau-t) + q_S(t)\delta(\tau-t) \right] \nn \\ &= \tau^2q_S(\tau) - \int_0^{\tau} dt~ t^2\frac{\partial q_S(t)}{\partial t} \nn \\ &= \tau^2q_S(\tau) -\tau^2q_S(\tau) + 2\int_0^{\tau} dt~ t~q_S(t) \nn \\ &= 2\int_0^{\tau} dt~ t~q_S(t).
\label{squared min}
\end{align}
We now turn to calculate $\langle R_\text{min} \rangle$. Since $R$ is deterministic, $\langle R_\text{min} \rangle$ is simply $\tau$. This can also be explicitly shown using Eq.(\ref{rmin-tmin-PDF})
\begin{align}
\langle R_\text{min} \rangle&=\int_0^{\infty} dt~ tf_{R_{min}}(t) \nn \\ &= \frac{1}{\text{Pr}(\tau\leq S)}\int_0^{\infty} dt~ t\delta(t-\tau) \text{Pr}(S>t) \nn \\ &= \frac{\tau q_S(\tau)}{q_S(\tau)} = \tau.
\label{Sharp Rmin}
\end{align}

Substituting Eqs. (\ref{mean min}), (\ref{Sharp denominator}), (\ref{squared min}) and (\ref{Sharp Rmin}) into \eref{second moment Sr} we arrive at the following expression
\begin{align}
\langle S_\tau^2  \rangle&=\frac{\langle \text{min}(S,\tau)^2 \rangle}{\text{Pr}(S<\tau)}+
\frac{2 \text{Pr}(\tau \leq S) \langle   R_{\text{min}}\rangle \langle \text{min}(S,\tau) \rangle}{\text{Pr}(S<\tau)^2} \nn \\ &= \frac{2\int_0^{\tau} dt~ tq_S(t)}{1-q_S(\tau)}+
\frac{2 \tau q_S(\tau)  \int_0^{\tau} dt~ q_S(t)}{(1-q_S(\tau))^2} \nn \\
&= \frac{2(1-q_S(\tau))\int_0^{\tau} dt~ tq_S(t)+2\tau q_S(\tau) \int_0^{\tau} dt~ q_S(t)}{(1-q_S(\tau))^2},
\end{align}
which was used in the main text (see \eref{second moment sharp restart}).


\end{document}